\documentclass[slac_one]{revtex4}
\usepackage{graphicx}
\usepackage{fancyhdr}
\begin{document}

\title{New QCD Results from the H1 Experiment at HERA} 

%

\author{B.Antunovic (for the H1 collaboration)}
\affiliation{Deutsches Elektronen Synchrotron, Notkestrasse 85, D-22603 Hamburg, Germany}

\begin{abstract}
New QCD results obtained by the H1 Collaboration are summarized. These results are based on data taken in $e^{\pm}p$ collisions at HERA~I (1994-2002) and HERA~II (2003-2007) corresponding to an integrated luminosity of almost $\rm 0.5~fb^{-1}$. The data were taken at an electron\footnote{Electron means electron and positron.} beam energy of $E_{e}=27.5$~GeV and proton beam energies of $\rm E_{p}=820~GeV, 920~GeV, 575~GeV$ and $\rm 460~GeV$. A direct measurement of $F_{L}$, jet production and measurement of $\alpha_{s}$, production of heavy quarks and diffractive photo-production of jets are presented.
\end{abstract}

\maketitle

\thispagestyle{fancy}


\section{MEASUREMENT OF THE PROTON STRUCTURE FUNCTION {\boldmath $F_{L}$}} 

\begin{figure}[hbt]
  \hfill
  \begin{minipage}[t]{.45\textwidth}
    \begin{center}
\hspace{-1.2cm}
\includegraphics[width=70mm]{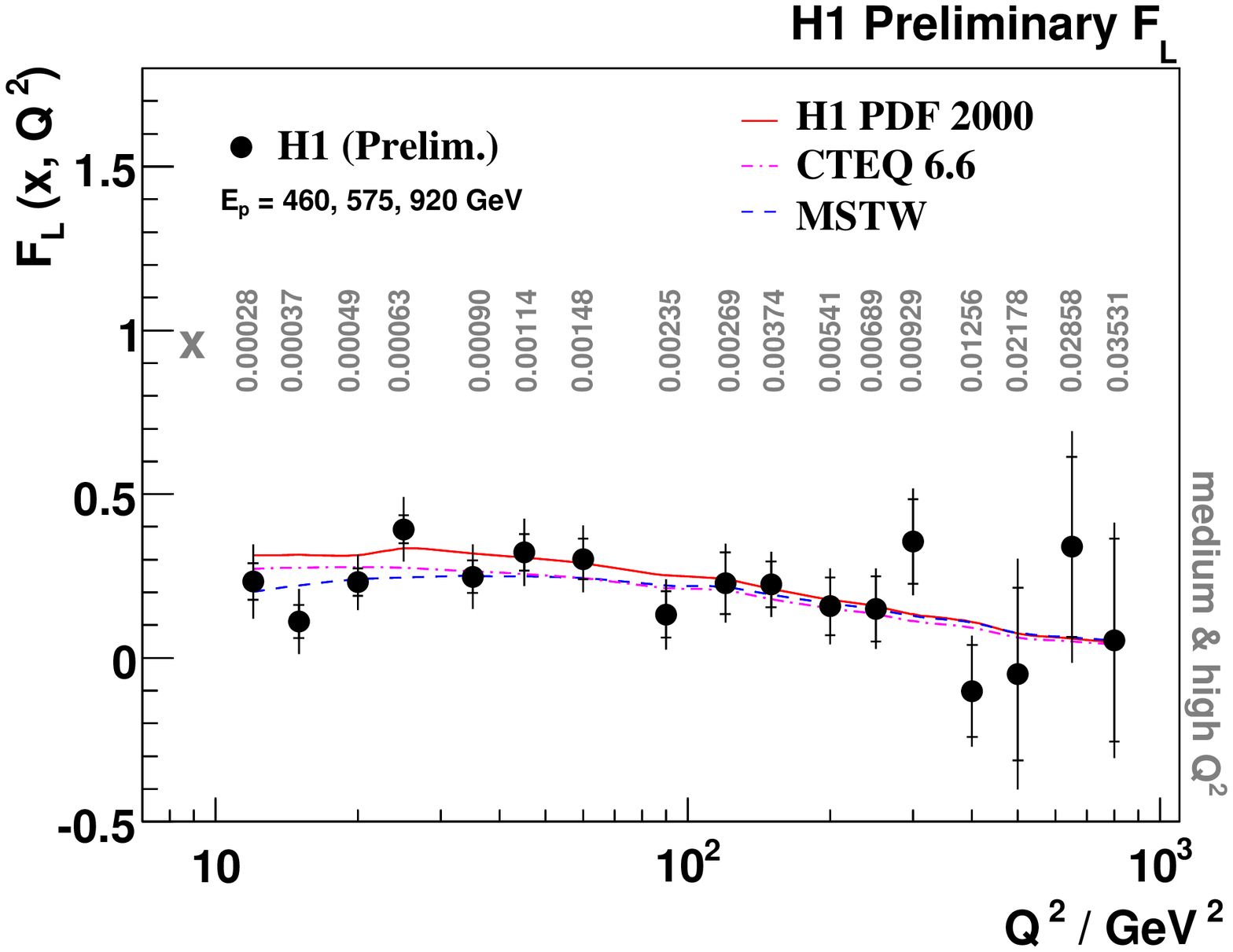}
\caption{The proton structure function $F_{L}$ measured as a function of $Q^{2}$ for given values of $x$ (indicated in grey). The solid line curve describes the expectation on $F_{L}(Q^{2},x)$ for the H1PDF 2000 using NLO QCD fit. The dashed line (dashed-dotted) is the expectation of the MSTW\cite{mstw} (CTEQ\cite{cteq}) pdf set using NNLO (NLO) QCD.} \label{BApic1} 
    \end{center}
  \end{minipage}
  \hfill
\begin{minipage}[t]{.45\textwidth}
    \begin{center}
\vspace{-6.5cm}
\includegraphics[width=60mm, angle=270]{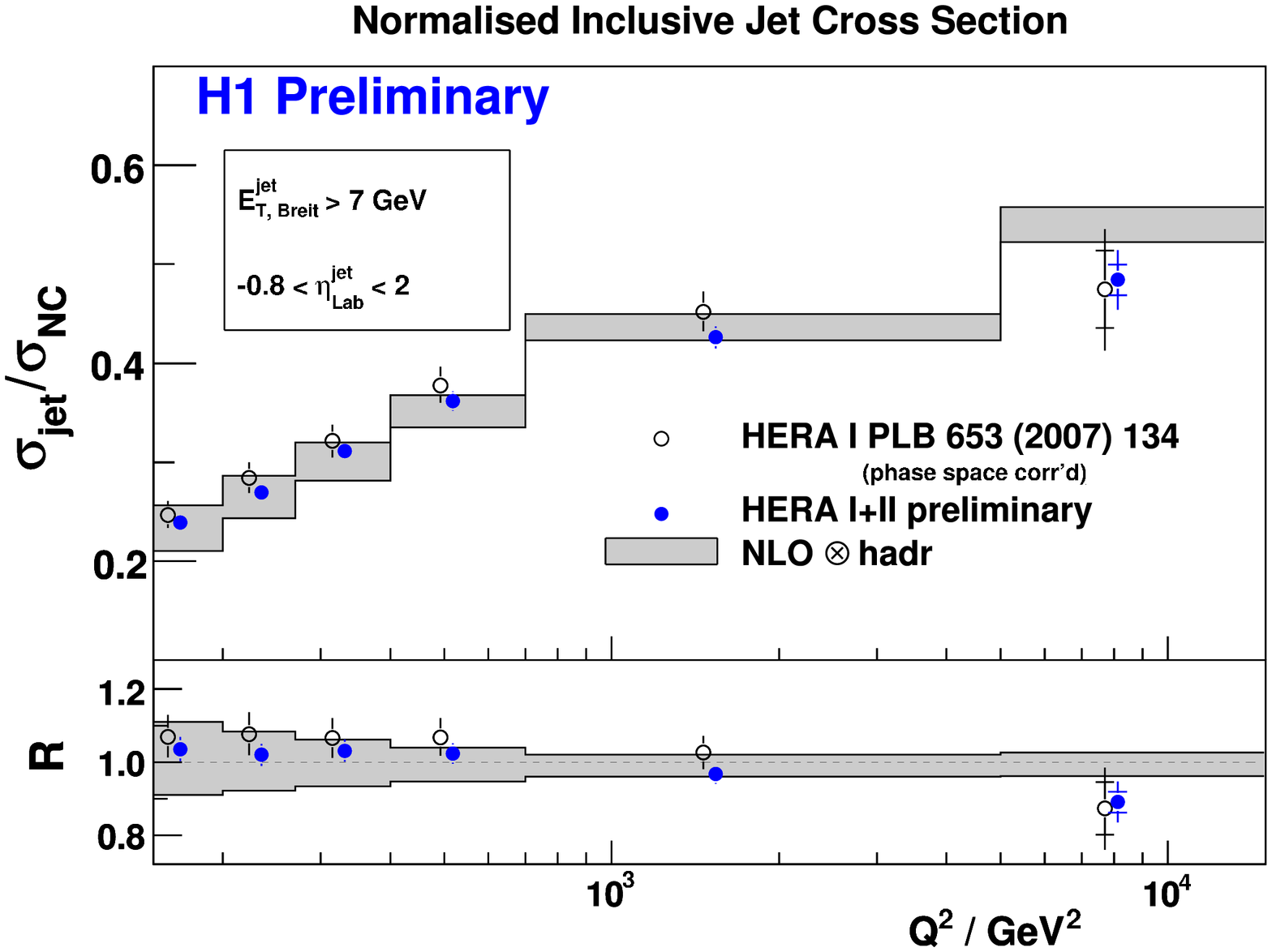}
\vspace{+0.1cm}
\caption{Ratio of the inclusive jet to the NC cross section as a function of $Q^{2}$ measured using HERA~I+II data compared to the HERA~I published data. The NLO QCD prediction including the theory uncertainties are shown as grey bands. The ratio $R$ of data with respect to the NLO QCD prediction is shown in the lower plot.} \label{BApic2}
    \end{center}
  \end{minipage}
  \hfill
\end{figure}

The first direct measurement of the proton structure function $F_{L}(x, Q^{2})$ performed by the H1 collaboration at HERA is presented. It is based on the measurement of the inclusive neutral current (NC) deep inelastic scattering (DIS) cross section which is given at low $Q^{2}$ by:
\begin{equation}\label{eq:redxsec}
\sigma_{r}(x,Q^2, y) = \frac{d^{2}\sigma}{dxdQ^{2}}  \cdot \frac{Q^{4}x}{2\pi \alpha^{2} Y_{+}}=F_{2}(x,Q^{2})-\frac{y^{2}}{Y_{+}}F_{L}(x,Q^{2})\;,\;\;\; Y_{+}=1+(1-y)^{2}\;.
\end{equation}
Here $x$ is the Bjorken variable which in the Quark Parton Model (QPM) corresponds to the fraction of the proton momentum carried by the struck quark, $Q^{2}$ is the negative four-momentum transfer squared, $y=Q^{2}/sx$ is the inelasticity and $s=4E_{p}E_{e}$ is the center of mass energy squared. The direct measurement of the proton structure function $F_{L}$ at HERA is based on the measurement of the reduced cross section (see Eq.~\ref{eq:redxsec}) at the same values of $x$ and $Q^{2}$ by varying the inelasticity $y$. This is achieved by changing the proton beam energy $\rm E_{p}=920, 575, 460$~GeV while keeping the positron beam energy of $\rm E_{e}=27.5$~GeV constant which leads to a variation of the center of mass energy. The reduced DIS cross section $\sigma_{r}(x,Q^{2},y)$ is measured in the range $\rm 12 \le Q^{2} \le 800~GeV^{2}$ and $0.00024 \le x \le 0.0015$. The structure function $F_{L}$ is determined from linear fits to the $\sigma_{r}(x,Q^{2},y)$ as a function of $y^{2}/Y_{+}$ using statistical and systematic uncertainties. The measurements of the structure function $F_{L}$ are performed as function of $x$ and $Q^{2}$ as shown in Fig.~\ref{BApic1}. The results are in good agreement with the expectations from the perturbative QCD (pQCD) confirming a dominant gluon contribution at low $x$.

\section{INCLUSIVE AND MULTI-JET PRODUCTION}
\label{sec:hfs}
Jet production is directly sensitive to $\alpha_{s}$ and allows for a precision test of QCD. The inclusive jet, 2-jet and 3-jet cross sections, normalized to the NC DIS cross section are measured in the range of $\rm 5<Q^{2}<100~GeV^{2}$ and $\rm 150<Q^{2}<15000~GeV^{2}$ by the H1 Collaboration. Particles of the hadronic final state are clustered into jets using the inclusive $k_{T}$ algorithm with the $p_{T}$ recombination scheme and with the distance parameter $R=1$ in the $\eta_{B}-\phi_{B}$ plane. The normalized inclusive cross sections are compared with the NLO calculations and recently published measurements by H1 based on HERA~I data. The strong coupling $\alpha_{s}$ is extracted from fits of the NLO prediction to the individual jet measurements as well as their combination. The experimental uncertainty of $\alpha_{s}$ is defined by the change in $\alpha_{s}$ which gives an increase of one unit in $\chi^{2}$ with respect to the minimal value. The dominating source for the theoretical error is the renormalization scale dependence reflecting the effect of missing orders beyond NLO in the pQCD prediction. The strong coupling is extracted from the combined jet measurements as $\rm \alpha_{s}(\mu_{z})=0.1182\pm 0.0008(exp.)\pm^{0.0041}_{0.0031}(th.) \pm 0.018(pdf.)$.
\section{DIFFRACTIVE PHOTO-PRODUCTION OF JETS}
\label{sec:diff}

\begin{figure*}[hbt]
\centering
\includegraphics[width=70mm]{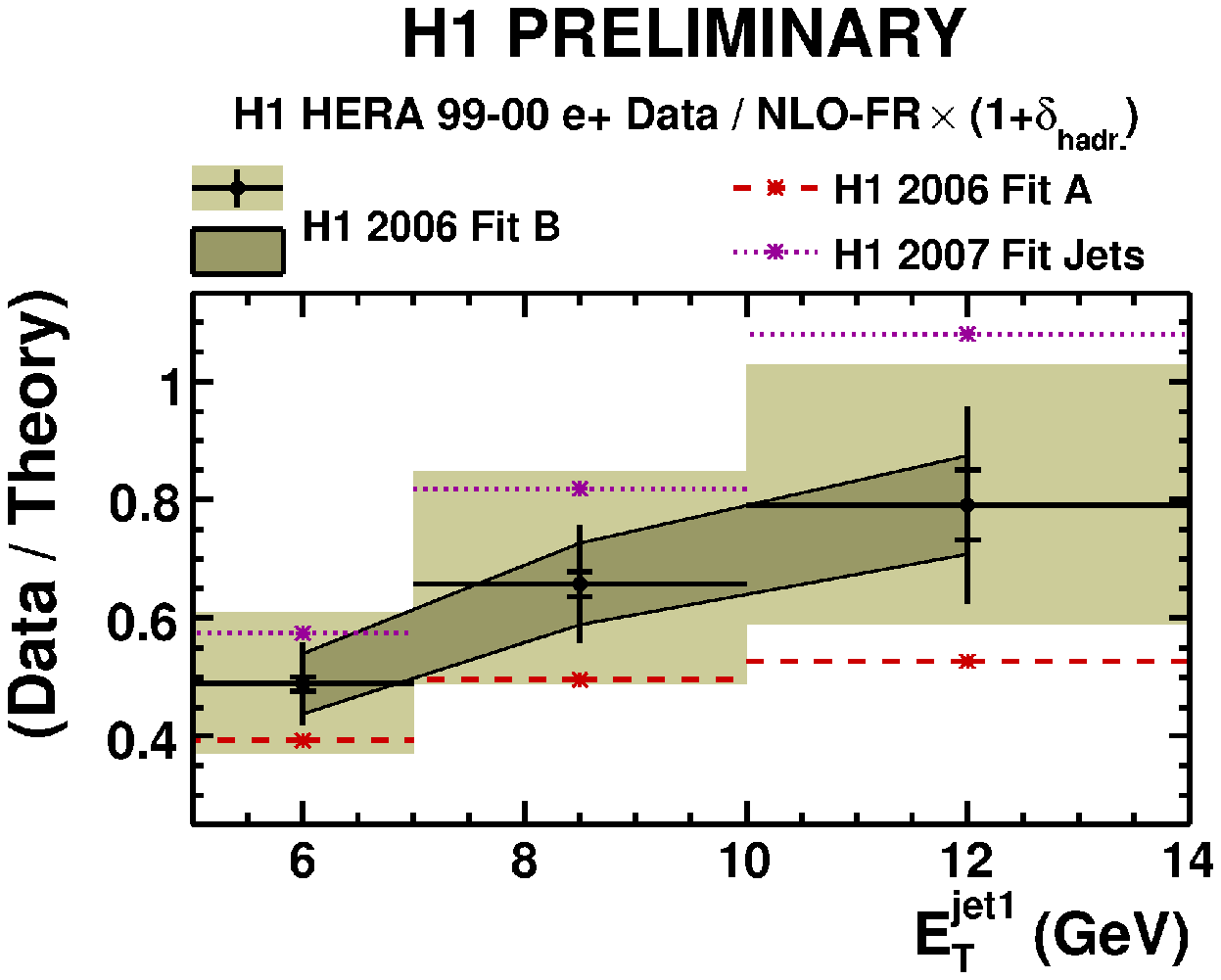}
\includegraphics[width=70mm]{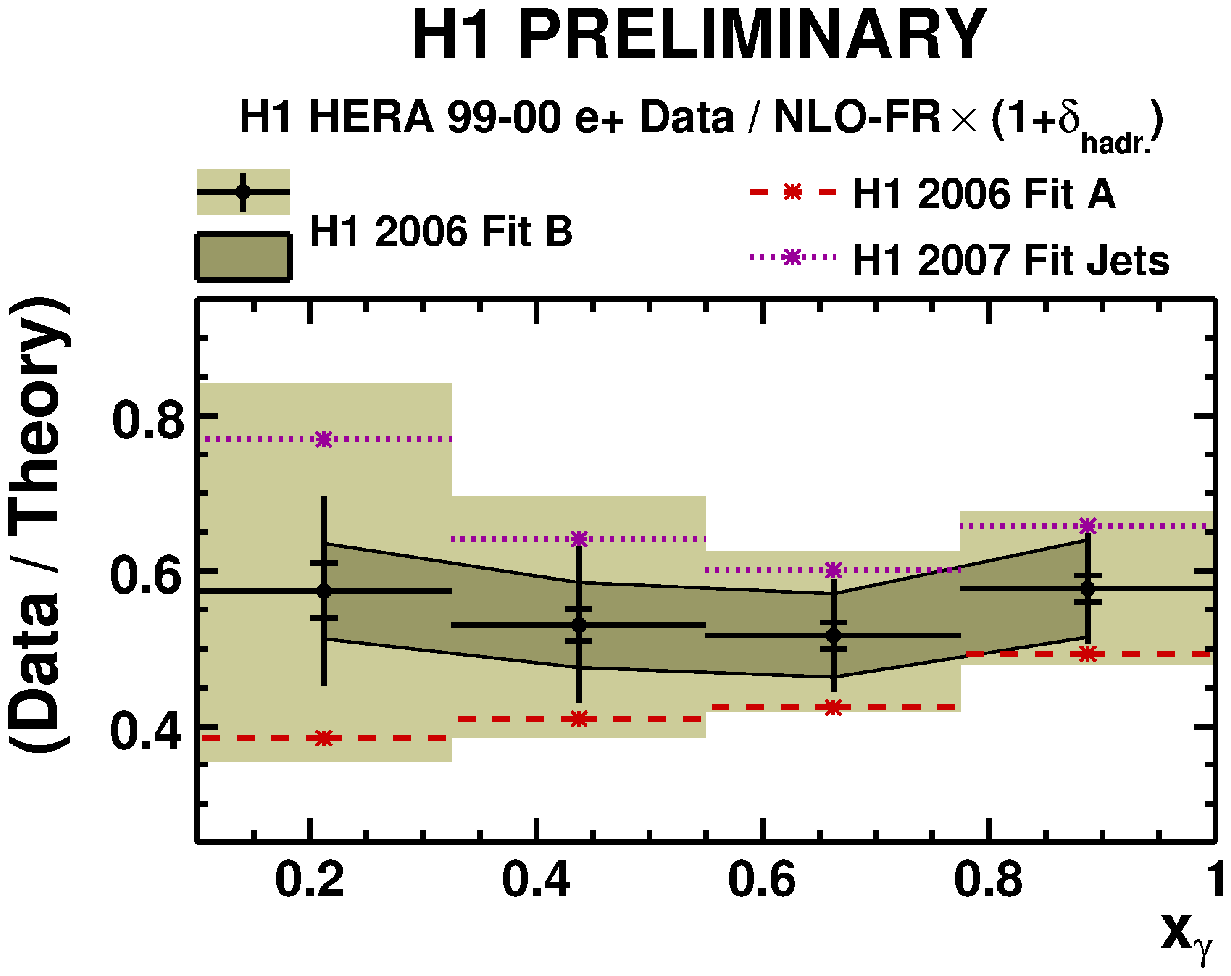}
\caption{Ratio of the measured differential cross section to the FR\cite{frixione} calculation based on the H1 2006 fit B DPDF set and corrected for hadronisation effects. The experimental and theory scale uncertainties are shown as bands.} \label{BApic5}
\end{figure*}

 The differential dijet cross sections in diffractive photo-production $\gamma p \rightarrow \gamma XY$ ($\rm Q^{2}<0.01~GeV^{2}$) have been measured. The event topology is given by the system $X$, containing at least two jets, separated from a leading low-mass proton dissociative system $Y$ by large rapidity gap. If factorization holds the process can be described by the QCD calculable hard subprocess and diffractive parton distribution functions (DPDF). This factorization can be broken in resolved processes, where the additional photon remnant and proton are expected to fill the large rapidity gap and destroy the experimental diffractive signature. The measurements are performed in two kinematic ranges differing in the transverse energy requirements on the two hardest jets. The diffractive dijet events for both direct and resolved processes are compared to the expectation in Fig.~\ref{BApic5}, where the measurements are shown normalized to the cross sections. The NLO calculations predict larger cross sections than the data. The suppression of the data relative to calculation is found not to have a significant dependence on the photon four-momentum fraction $x_{\gamma}$ entering the hard subprocess, while a dependence of the suppression factor on the transverse energy of the jets is suggested.

\section{\boldmath{$D^{*\pm}$} PRODUCTION}
\label{sec:hf}

\begin{figure}[h]
  \hfill
  \begin{minipage}[t]{.45\textwidth}
    \begin{center}
\vspace{1cm}
\includegraphics[width=70mm]{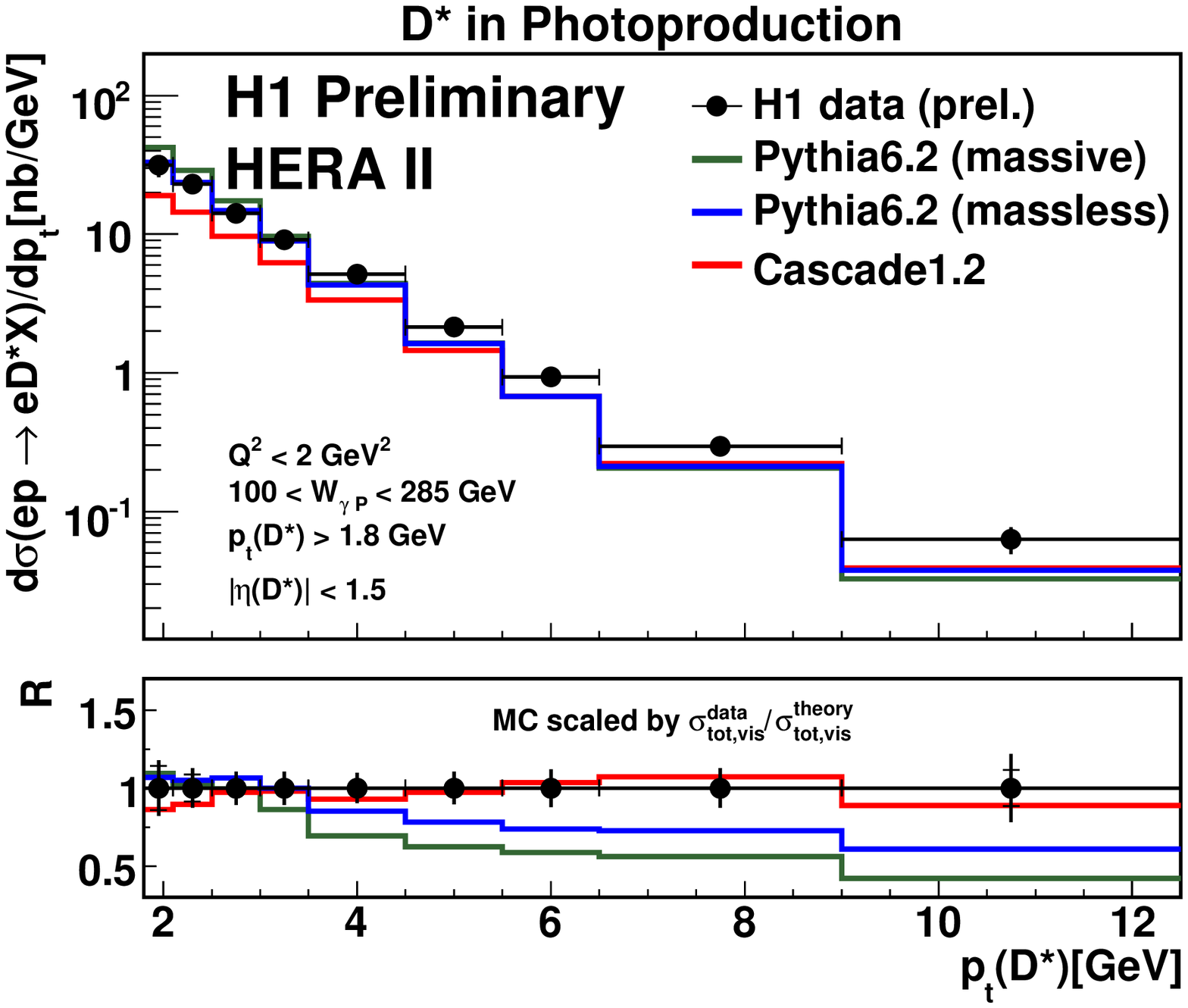}
\vspace{-2cm}
\caption{The differential cross section as a function of $p_{t}(D^{*})$ measured by H1. The green (blue) histogram shows the prediction of MC program Pythia 6.2 where the charm quarks are treated as the massive (massless) particles. The red histogram shows the prediction of the Cascade 1.2.} \label{BApic3} 
    \end{center}
  \end{minipage}
  \hfill
\begin{minipage}[t]{.45\textwidth}
    \begin{center}
\vspace{+1.4cm}
\hspace{-1.2cm}
\includegraphics[width=60mm]{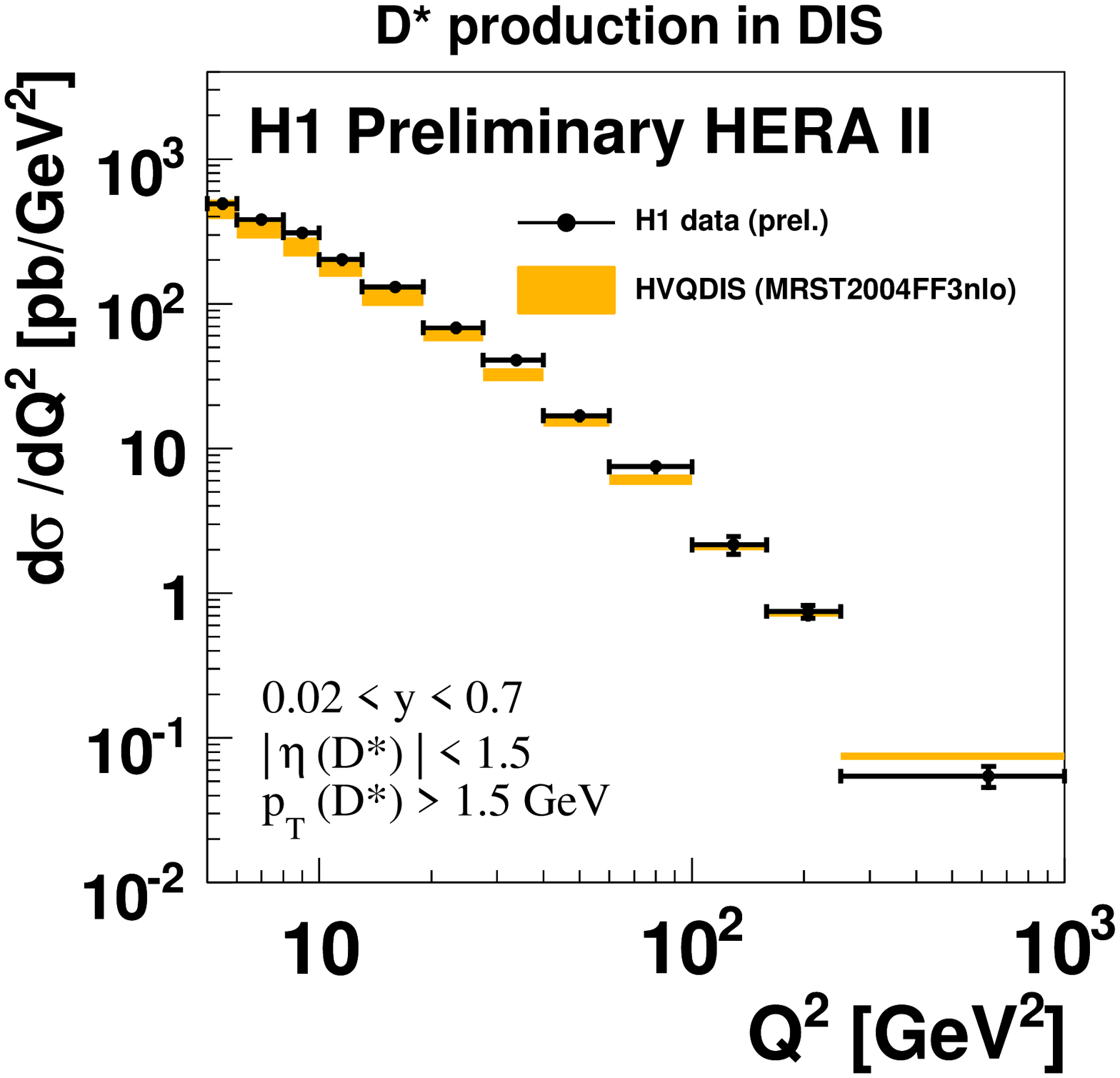}
\vspace{+0.1cm}
\caption{The differential cross sections as a function of $Q^{2}$ compared with the NLO calculation HVQDIS.} \label{BApic4}
    \end{center}
  \end{minipage}
  \hfill
\end{figure}

 The inclusive production of $D^{*\pm}$ mesons has been measured in the kinematic region of small photon virtuality $\rm Q^{2}<2~GeV^{2}$ (photo-production) and high photon virtuality $\rm 5< Q^{2}<100~GeV^{2}$ (DIS). The dominant heavy flavor production process is photon gluon-fusion ($\gamma g \rightarrow Q \bar{Q}$). Charm events are tagged via full reconstruction using the decay chain $D^{* \pm} \rightarrow D^{0}+\pi^{\pm}_{slow} \rightarrow K^{\mp}+\pi^{\pm}+\pi^{\pm}_{slow}$. The $D^{*\pm}$ production measured in the photo-production events is compared to the prediction of CASCADE 1.2\cite{cascade1,cascade2} and PYTHIA 6.2\cite{pythia} where the charm quarks are treated as massive (massless) particles. CASCADE 1.2 describes the shape of $p_t(D^{*})$, but is too low in normalization. The expectations of PYTHIA 6.2 are too steep as can be seen in Fig~\ref{BApic3}. The $D^{*\pm}$ production in the DIS events is compared to the NLO calculation HVQDIS and shows remarkably good agreement as shown in Fig.~\ref{BApic4}.
\vspace{-1.2cm}
\vspace{-0.9cm}
\section{SUMMARY}
\label{sec:summ}
A selection of recent results on QCD processes from the H1 experiment at HERA has been presented: the first direct measurement of the longitudinal proton structure function at low $x$, the strong coupling $\alpha_{s}$ measured with a high accuracy competitive with the world average value, $D^{*}$ production in photo-production and DIS and diffractive dijet production. The results give new insights on the proton structure and electron-proton interactions, reaching the era of impressive high precision at HERA.
\vspace{-0.5cm}
%


\end{document}